\documentstyle[12pt]{article}
\newlength{\extraspace}
\newlength{\extraspaces}
\newcommand{\be}{\begin{equation}
\addtolength{\abovedisplayskip}{\extraspaces}
\addtolength{\belowdisplayskip}{\extraspaces}
\addtolength{\abovedisplayshortskip}{\extraspace}
\addtolength{\belowdisplayshortskip}{\extraspace}}
\newcommand{\ee}{\end{equation}}
\newcommand{\ba}{\begin{eqnarray}
\addtolength{\abovedisplayskip}{\extraspaces}
\addtolength{\belowdisplayskip}{\extraspaces}
\addtolength{\abovedisplayshortskip}{\extraspace}
\addtolength{\belowdisplayshortskip}{\extraspace}}
\newcommand{\ea}{\end{eqnarray}}
\newcommand{\nonu}{\nonumber \\[.5mm]}
\newcommand{\A}{&\!\!\!}
\begin{document}
\thispagestyle{empty}
\setlength{\baselineskip}{6mm}
\begin{flushright}
SIT-LP-13/01 \\
%{\tt hep-th/} \\
January, 2013
\end{flushright}
\vspace{7mm}
\begin{center}
{\large\bf Composite representations of new SUSY algebra \\[2mm]
under nonlinear SUSY transformations in curved spacetime
} \\[20mm]
{\sc Kazunari Shima}
\footnote{
\tt e-mail: shima@sit.ac.jp} \ 
and \ 
{\sc Motomu Tsuda}
\footnote{
\tt e-mail: tsuda@sit.ac.jp} 
\\[5mm]
{\it Laboratory of Physics, 
Saitama Institute of Technology \\
Fukaya, Saitama 369-0293, Japan} \\[20mm]
\begin{abstract}
We discuss composite representations of new supersymmetry (SUSY) algebra 
under nonlinear (NL) SUSY transformations in curved space-time 
from the viewpoint of nonlinear supersymmetric general relativity (NLSUSY GR)/ SGM scenario. 
We show in the linearization of NLSUSY how various basic fields in linear SUSY theories/supergravity 
are expressed as functionals in terms of a vierbein and Nambu-Goldstone fermion fundamental fields 
in NLSUSY GR. 
%
%Supersymmetry (SUSY) algebra is automatically localized by means of a new global nonlinear (NL) SUSY 
%in NLSUSY general relativity (NLSUSYGR). 
%We discuss how various basic fields of standard model are expressed as representation of the NLSUSY algebra, 
%i.e. as composites of a Nambu-Goldstone spinor fields and a veirbein in curved spacetime. 
%
\\[5mm]
%
%\noindent
%PACS: 11.30.Pb, 12.60.Jv, 12.60.Rc, 12.10.-g \\[2mm]
%%
%\noindent
%Keywords: supersymmetry, superfield, composite unified theory 
%
\end{abstract}
\end{center}

\newpage

\noindent
Supersymmetry (SUSY) \cite{WZ} is a fundamental and promissing notion towards a unified field theories 
of space-time and matter beyond the standard model (SM). 
Various basic fields in linear (L) SUSY theories are representations of SUSY algebra 
where LSUSY transformations satisfy the closed commutator SUSY algebra. 
The SUSY algebra is localized (gauged) in supergravity (SUGRA) \cite{FNF,DZ} 
by means of the spin-connection formalism used in general relativity (GR) coupled with fermions. 

On the other hand, the localization of SUSY algebra is realized \cite{KS1,ST1} 
in nonlinear supersymmetric general relativity (NLSUSYGR) \cite{KS1} 
by introducing new nonlinear (NL) SUSY transformations for a vierbein 
and Nambu-Goldstone (NG) fermion fields in curved space-time. 
NLSUSYGR was constructed based on the GR principle and NLSUSY \cite{VA} as a representation 
of $SO(10)$ super-Poincar\'e (SP) algebra which minimally accomodates all observed particles in SM 
and the graviton \cite{KS1,KS2}. 
The fundamental action of NLSUSYGR is given as an Einstein-Hilbert (EH) form with a cosmological term 
by defining a {\it unified} vierbein in a (unstable) space-time whose tangent space-time 
are denoted by $SL(2,{\bf C})$ coset (super$GL(4,{\bf R})$/$GL(4,{\bf R})$) 
Grassmann coordinates in addition to $SO(3,1)$ Minkowski ones. 
Corresponding to the spontaneous breakdown of the space-time in NLSUSYGR to the Riemann space-time 
with (massless) NG fermions, 
NLSUSYGR is described by means of the EH action of GR and the highly nonlinear interactions 
of the ordinary vierbein (graviton) and the NG fermions. 

Based on the (localized) SUSY algebra, NLSUSYGR and LSUSY theories / SUGRA 
would relate with each other through the linearization of NLSUSY, 
where various basic fields in LSUSY theories/SUGRA 
are expressed as functionals (composites) of an ordinary vierbein and the NG fermions in NLSUSYGR. 
In flat space-time, the relation between Volkov-Akulov (VA) NLSUSY model \cite{VA} 
and various LSUSY theories (NL/LSUSY relation) are systematically obtained 
by means of the linearization of NLSUSY \cite{IK}-\cite{UZ} and we have shown the NL/LSUSY relation 
in more realistic $N = 2$ SUSY (QED and Yang-Mills) theories in flat space-time \cite{STT}-\cite{ST4-lin}. 
Since the action in NLSUSYGR reduces to the VA NLSUSY one in flat space-time, 
the NLSUSYGR/SGM scenario \cite{KS1,ST1,KS2,ST5-SGM} gives new insights 
into the origin of mass and (bare) gauge coupling constant from the NL/LSUSY relation \cite{STL,ST6-gauge}. 
We have also discussed the linearization of NLSUSY in curved space-time 
for a relation between the (most) general supermultiplet of $N = 1$ SUGRA 
and the fundamental fields in NLSUSYGR \cite{STS,ST7}. 

In order to develop the argument of NL/LSUSY relation in curved space-time, 
we discuss in this letter the new SUSY algebra under the NLSUSY transformations in NLSUSYGR 
and the composite representations for various basic fields of LSUSY theories/SUGRA 
in terms of the vierbein and NG fermion fields. 
We show that the basic fields in LSUSY theories/SUGRA have to be represented as functionals of the vierbein, 
the NG fermions and their (anticommutative) first-order derivatives in the linearization of NLSUSY. 

For self-contained arguments, let us briefly review a fundamental action of NLSUSYGR 
and NLSUSY transformations in curved space-time \cite{KS1}. 
The space-time structure of NLSUSYGR is a curved one whose tangent space-time is denoted 
by means of $SO(3,1)$ Minkowski and $SL(2,{\bf C})$ Grassmann (NG fermion) coordinates $(x^a, \psi^i)$. 
A unified vierbein $w^a{}_{\mu}$ is defined as 
\be
w^a{}_\mu = e^a_\mu + t^a{}_\mu, 
\ee
where $e^a_\mu$ is an ordinary vierbein in Riemann space-time 
and $t^a{}_\mu = - i \kappa^2 \bar\psi^i \gamma^a \partial_\mu \psi^i$ with a constant $\kappa$ 
whose dimension is (length)$^2 =$ (mass)$^{-2}$. 
The action (Lagrangian density) in NLSUSYGR is expressed as an EH form with a cosmological term, 
\be
{\cal L} = -{c^4 \over 16 \pi G} \vert w \vert (\Omega + \Lambda), 
\label{action}
\ee
where $\vert w \vert = \det w^a{}_\mu$, $\Omega$ is a scalar curvature in terms of ($w^a{}_\mu$, $w_a{}^\mu$) 
and $\Lambda$ means a cosmological constant. 
In the (Riemann-)flat space-time case ($e^a_\mu \rightarrow \delta^a_\mu$), the action (\ref{action}) 
reduces to a VA NLSUSY action, 
\be
L = -{1 \over 2 \kappa^2} \vert w \vert 
= -{1 \over 2 \kappa^2} \left\{ 1 + t^a{}_a + {1 \over 2}(t^a{}_a t^b{}_b - t^a{}_b t^b{}_a) 
+ \cdots \right\}, 
\ee
where the constant $\kappa$ is fixed as $\displaystyle{\kappa^{-2} = {c^4 \Lambda \over 8 \pi G}}$. 

The NLSUSYGR action (\ref{action}) is invariant under the following (new) NLSUSY transformations 
of the NG fermions $\psi^i$ and the vierbein $e^a_\mu$, 
\ba
\delta_\zeta \psi^i \A = \A {1 \over \kappa} \zeta^i - i \kappa \bar\zeta^j \gamma^\mu \psi^j \partial_\mu \psi^i, 
\nonu
\delta_\zeta e^a_\mu \A = \A 2i \kappa \bar\zeta^i \gamma^\nu \psi^i \partial_{[\mu} e^a_{\nu]}, 
\label{NLSUSY}
\ea
with global spinor parameters $\zeta^i$, which induces $GL(4,{\bf R})$ transformations of $w^a{}_\mu$, 
\be
\delta_\zeta w^a{}_\mu 
= \xi^\nu \partial_\nu w^a{}_\mu + w^a{}_\nu \partial_\mu \xi^\nu, 
\ee
where $\xi^\mu = -i \kappa \bar\zeta^i \gamma^\mu \psi^i$. 
The NLSUSY transformations (\ref{NLSUSY}) satisfy a closed 
(and localized) commutator algebra to $GL(4,{\bf R})$ \cite{ST1}, 
\be
[\delta_{\zeta_1}, \delta_{\zeta_2}] = \delta_{GL(4,{\bf R})}(\Xi^\mu) 
\label{algebra}
\ee
with a $GL(4,{\bf R})$ transformation parameters 
\be
\Xi^\mu = 2(i \bar\zeta_1^i \gamma^\mu \zeta_2^i - \xi_1^\nu \xi_2^\rho e_a^\mu \partial_{[\nu} e^a_{\rho]}). 
\ee
The NLSUSY GR action (\ref{action}) possesses (promissing) large symmetries 
accomodating $SO(N)$ ($SO(10)$) SP group, namely, they are invariant under \cite{KS1,ST1,ST5-SGM} 
\ba
\A \A 
[{\rm new \ NLSUSY}] \otimes [{\rm local}\ GL(4,{\bf R})] \otimes [{\rm local \ Lorentz}] 
\nonu
\A \A 
\otimes [{\rm local \ spinor \ translation}] 
\otimes [{\rm global}\ SO(N)] \otimes [{\rm local}\ U(1)^N] \otimes [{\rm chiral}]. 
\ea

Let us discuss below possible {\it composite} representations of the new SUSY algebra (\ref{algebra}) 
under the NLSUSY transformations (\ref{NLSUSY}). 

In flat space-time, we assume that various basic fields in LSUSY theories are expressed as functionals 
of $\psi^i$ and its first- and higher-order derivatives ($\partial \psi^i$, $\partial^2 \psi^i$, $\cdots$) 
in the linearization of NLSUSY as, 
\be
f_A = f_A(\psi^i, \partial \psi^i, \partial^2 \psi^i, \cdots), 
\ \ \ g_B = g_B(\psi^i, \partial \psi^i, \partial^2 \psi^i, \cdots) 
\ \ \ (A, B = a, ab, \cdots, {\rm etc.}) 
\label{functionals}
\ee
which satsify the familiar closed commutator algebras under the NLSUSY transformations of $\psi^i$, 
\be
[\delta_{\zeta_1}, \delta_{\zeta_2}] f_A = \Xi^a \partial_a f_A, 
\ \ \ [\delta_{\zeta_1}, \delta_{\zeta_2}] g_B = \Xi^a \partial_a g_B, 
\ee
with translational parameters $\Xi^a = 2i \bar\zeta_1^i \gamma^a \zeta_2^i$. 
Note that $f_A$ and $g_B$ include the case of spinor functionals. 
Then, commutator algebras for the products $f_A g_B$ are also closed as 
\be
[\delta_{\zeta_1}, \delta_{\zeta_2}] (f_A g_B) = \Xi^a \partial_a (f_A g_B). 
\ee
This means that all functionals in terms of ($\psi^i, \partial \psi^i, \partial^2 \psi^i, \cdots$) 
are the representations of SUSY algebra, $[\delta_{\zeta_1}, \delta_{\zeta_2}] = \delta_P(\Xi^a)$, 
since NLSUSY transformations of ($\psi^i$, $\partial \psi^i$, $\partial^2 \psi^i$, $\cdots$) 
satisfy the SUSY algebra, respectively. 

In the curved space-time case, let us first consider functionals of $\psi^i$ and $e^a{}_\mu$, 
\be
f^A{}_M = f^A{}_M(\psi^i, e^a{}_\mu), 
\ \ \ g^B{}_N = g^B{}_N(\psi^i, e^a{}_\mu), 
\ \ \ (M, N = \mu, \mu \nu, \cdots) 
\ee
with satisfying the NLSUSY algebra (\ref{algebra}) as 
\ba
\A \A 
[\delta_{\zeta_1}, \delta_{\zeta_2}] f^A{}_M = \Xi^\rho \partial_\rho f^A{}_M 
+ \sum_k f^A_{N_k(\rho)} \partial_{\mu_k} \Xi^\rho, 
\nonu
\A \A 
[\delta_{\zeta_1}, \delta_{\zeta_2}] g^A{}_M = \Xi^\rho \partial_\rho g^A{}_M 
+ \sum_k g^A_{N_k(\rho)} \partial_{\mu_k} \Xi^\rho. 
\label{algebra2}
\ea
In Eq.(\ref{algebra2}), $N_k(\rho) = \mu_1 \mu_2 \cdots \rho \cdots \mu_n$ 
and the summation $\displaystyle{\sum_k}$ means 
\ba
\sum_k f^A_{N_k(\rho)} \partial_{\mu_k} \Xi^\rho 
\A = \A f^A_{N_1(\rho)} \partial_{\mu_1} \Xi^\rho + f^A_{N_2(\rho)} \partial_{\mu_2} \Xi^\rho + \cdots 
\nonu
\A = \A f^A_{\rho \mu_2 \cdots} \partial_{\mu_1} \Xi^\rho + f^A_{\mu_1 \rho \cdots} \partial_{\mu_2} \Xi^\rho + \cdots. 
\ea
Then, the commutator algebras for the products $f^A{}_M g^B{}_N$ are also closed to $GL(4,{\bf R})$ as 
\ba
[\delta_{\zeta_1}, \delta_{\zeta_2}] (f^A{}_M g^B{}_N) 
\A = \A \Xi^\rho \partial_\rho (f^A{}_M g^B{}_N) 
+ \sum_k f^A_{L_k(\rho)} \partial_{\mu_k} \Xi^\rho g^B{}_N 
+ f^A{}_M \sum_k g^B_{L_k(\rho)} \partial_{\mu_k} \Xi^\rho 
\nonu
\A = \A \delta_{GL(4,{\bf R})} (f^A{}_M g^B{}_N)(\Xi^\rho). 
\label{algebra3}
\ea
Since the NLSUSY transformations (\ref{NLSUSY}) for $\psi^i$ and $e^a{}_\mu$ satisfy the NLSUSY algebra (\ref{algebra}), 
all functionals in terms of $\psi^i$ and $e^a{}_\mu$ are the representations of the SUSY algebra (\ref{algebra}) 
from Eqs.(\ref{algebra2}) and (\ref{algebra3}). 

As for commutator algebras for the derivatives of $\psi^i$ and $e^a{}_\mu$ 
under the NLSUSY transformations (\ref{NLSUSY}), 
only for $\partial_\mu \psi^i$ and $\partial_{[\mu} e^a{}_{\nu]}$ are closed as 
\ba
[\delta_{\zeta_1}, \delta_{\zeta_2}] \partial_\mu \psi^i 
\A = \A \Xi^\nu \partial_\nu (\partial_\mu \psi^i) + (\partial_\nu \psi^i) \partial_\mu \Xi^\nu
= \delta_{GL(4,{\bf R})} (\partial_\mu \psi^i)(\Xi^\nu), 
\nonumber\\[2mm]
[\delta_{\zeta_1}, \delta_{\zeta_2}] \partial_{[\mu} e^a{}_{\nu]} 
\A = \A \Xi^\rho \partial_\rho (\partial_{[\mu} e^a{}_{\nu]}) 
+ (\partial_{[\rho} e^a{}_{\nu]}) \partial_\mu \Xi^\rho 
+ (\partial_{[\mu} e^a{}_{\rho]}) \partial_\nu \Xi^\rho 
\nonu
\A = \A \delta_{GL(4,{\bf R})} (\partial_{[\mu} e^a{}_{\nu]})(\Xi^\nu). 
\label{algebra4}
\ea
The commutator algebra for the higher-order derivative terms of $\psi^i$ and $e^a{}_\mu$, 
i.e. for ($\partial^2 \psi^i$, $\partial^2 e^a{}_\mu$, $\cdots$), are not closed to $GL(4,{\bf R})$. 
Therefore, we conclude that in the linearization of NLSUSY 
the composite representations of the SUSY algebra (\ref{algebra}) in curved space-time 
are given by means of the functionals (including the spinor ones), 
\be
f^A{}_M = f^A{}_M(\psi^i, e^a{}_\mu; \partial_\mu \psi^i, \partial_{[\mu} e^a{}_{\nu]}) 
\label{functionals2}
\ee
from the same arguments as Eqs.(\ref{algebra2}) and (\ref{algebra3}). 
Note that the spin connection and the scalar curvature, etc. in GR are also the case, 
for they are the functionals of $e^a{}_\mu$ and $\partial_{[\mu} e^a{}_{\nu]}$. 

We summarize our results as follows. 
We have discussed in this letter the possible composite representations of the new SUSY algebra (\ref{algebra}) 
under the NLSUSY transformations (\ref{NLSUSY}) in the linearization of NLSUSY. 
Based on the commutator algebra for the vierbein, the NG fermions 
and their (anticommutative) first-order derivatives in Eqs.(\ref{algebra}) and (\ref{algebra4}), 
we conclude the functionals (\ref{functionals2}) is the representations 
by means of the arguments (\ref{algebra2}) and (\ref{algebra3}). 
This is in contrast with the flat space-time case (\ref{functionals}) 
depending for all higher-order derivatives of $\psi^i$. 
The argument of this letter may give an important insight into the linearization of NLSUSY in curved space-time, 
i.e. the relation between NLSUSYGR and LSUSY theories/SUGRA \cite{STS,ST7}.

\newpage

%%%%%%%  References  %%%%%%%%%%%%%%%%%%%%%%%%%%%%%%%%%%%%%%%
%
\newcommand{\NP}[1]{{\it Nucl.\ Phys.\ }{\bf #1}}
\newcommand{\PL}[1]{{\it Phys.\ Lett.\ }{\bf #1}}
\newcommand{\CMP}[1]{{\it Commun.\ Math.\ Phys.\ }{\bf #1}}
\newcommand{\MPL}[1]{{\it Mod.\ Phys.\ Lett.\ }{\bf #1}}
\newcommand{\IJMP}[1]{{\it Int.\ J. Mod.\ Phys.\ }{\bf #1}}
\newcommand{\PR}[1]{{\it Phys.\ Rev.\ }{\bf #1}}
\newcommand{\PRL}[1]{{\it Phys.\ Rev.\ Lett.\ }{\bf #1}}
\newcommand{\PTP}[1]{{\it Prog.\ Theor.\ Phys.\ }{\bf #1}}
\newcommand{\PTPS}[1]{{\it Prog.\ Theor.\ Phys.\ Suppl.\ }{\bf #1}}
\newcommand{\AP}[1]{{\it Ann.\ Phys.\ }{\bf #1}}


\begin{thebibliography}{100}

\bibitem{WZ}
J. Wess and B. Zumino, {\it Phys. Lett. B} {\bf 49} (1974) 52. 

\bibitem{FNF}
D. Z. Freedman, P. van Nieumenhuisen and S. Ferrara, {\it Phys. Rev. D} {\bf 13} (1976) 3214. 

\bibitem{DZ}
S. Deser and B. Zumino, {\it Phys. Lett. B} {\bf 62} (1976) 335. 

\bibitem{KS1}
K. Shima, {\it Phys. Lett. B} {\bf 501} (2001) 237. 

\bibitem{ST1}
K. Shima and M. Tsuda, {\it Phys. Lett. B} {\bf 507} (2001) 260. 

\bibitem{VA}
D.V. Volkov and V.P. Akulov, {\it Phys. Lett. B} {\bf 46} (1973) 109. 

\bibitem{KS2}
K. Shima, {\it Z. Phys. C} {\bf 18} (1983) 25; \\
K. Shima, {\it European Phys. J. C} {\bf 7} (1999) 341. 

\bibitem{IK}
E.A. Ivanov and A.A. Kapustnikov, {\it J. Phys. A} {\bf 11} (1978) 2375. 

\bibitem{Ro}
M. Ro\v{c}ek, {\it Phys. Rev. Lett.} {\bf 41} (1978) 451. 

\bibitem{UZ}
T. Uematsu and C.K. Zachos, {\it Nucl. Phys. B} {\bf 201} (1982) 250. 

\bibitem{STT}
K. Shima, Y. Tanii and M. Tsuda, {\it Phys. Lett. B} {\bf 546} (2002) 162. 

\bibitem{ST2-lin}
K. Shima and M. Tsuda, {\it Phys. Lett. B} {\bf 666} (2008) 410. 

\bibitem{ST3-lin}
K. Shima and M. Tsuda, {\it Mod. Phys. Lett. A} {\bf 24} (2009) 185. 

\bibitem{ST4-lin}
K. Shima and M. Tsuda, {\it Phys. Lett. B} {\bf 687} (2010) 89. 

\bibitem{ST5-SGM}
K. Shima and M. Tsuda, {\it Class. Quant. Grav.} {\bf 19} (2002) 5101; \\
K. Shima and M. Tsuda, {\it PoS HEP2005} (2006) 011. 

\bibitem{STL}
K. Shima and M. Tsuda, {\it Phys. Lett. B} {\bf 645} (2007) 455; \\
K. Shima, M. Tsuda and W. Lang, {\it Phys. Lett. B} {\bf 659} (2008) 741. 

\bibitem{ST6-gauge}
K. Shima and M. Tsuda, {\em Nuovo Cim. B} {\bf 124} (2009) 645. 

\bibitem{STS}
K. Shima, M. Tsuda and M. Sawaguchi, {\it Int. J. Mod. Phys. E} {\bf 13} (2004) 539. 

\bibitem{ST7}
K. Shima and M. Tsuda, {\it Phys. Lett. B} {\bf 609} (2005) 385. 












%%%%%%%%%%%%%%%%%%%%(VA NLSUSY)
%
%\bibitem{VA}
%%D.V. Volkov and V.P. Akulov,  
%%{\it JETP Lett.} {\bf 16} (1972) 438; \\
%D.V. Volkov and V.P. Akulov, {\it Phys. Lett. B} {\bf 46} (1973) 109. 
%
%%%%%%%%%%%%%%%%%%%%(SUSY)
%
%\bibitem{GL}
%Y.A. Golfand and E.S. Likhtman, 
%{\it JET Lett.} {\bf 13} (1971) 323. 
%
%\bibitem{WZ1}
%J. Wess and B. Zumino, {\it Phys. Lett. B} {\bf 49} (1974) 52. 
%
%\bibitem{WZ2}
%J. Wess and B. Zumino, {\it Nucl. Phys. B} {\bf 78} (1974) 1. 
%
%%%%%%%%%%%%%%%%%%%%(SGM scenario)
%
%\bibitem{KS1}
%K. Shima, {\it Phys. Lett. B} {\bf 501} (2001) 237. 
%%K. Shima, {\it Z. Phys. C} {\bf 18} (1983) 25; \\
%
%\bibitem{KS2}
%K. Shima, {\it European Phys. J. C} {\bf 7} (1999) 341. 
%
%\bibitem{ST1-SGM}
%K. Shima and M. Tsuda, {\it Phys. Lett. B} {\bf 507} (2001) 260. 
%
%\bibitem{ST2-SGM}
%K. Shima and M. Tsuda, {\it Class. Quant. Grav.} {\bf 19} (2002) 5101. 
%
%\bibitem{ST3-SGM}
%K. Shima and M. Tsuda, {\it PoS HEP2005} (2006) 011. 
%
%%%%%%%%%%%%%%%%%%%%(linearization of NLSUSY and NL/L SUSY relation for free theories)
%
%\bibitem{IK1}
%%E.A. Ivanov and A.A. Kapustnikov, 
%%Relation between linear and nonlinear realizations of 
%%supersymmetry, JINR Dubna Report No. E2-10765, 1977 (unpublished); \\
%E.A. Ivanov and A.A. Kapustnikov, {\it J. Phys. A} {\bf 11} (1978) 2375. 
%
%\bibitem{IK2}
%E.A. Ivanov and A.A. Kapustnikov, {\it J. Phys. G} {\bf 8} (1982) 167. 
%
%\bibitem{Ro}
%M. Ro\v{c}ek, {\it Phys. Rev. Lett.} {\bf 41} (1978) 451. 
%
%\bibitem{UZ}
%T. Uematsu and C.K. Zachos, {\it Nucl. Phys. B} {\bf 201} (1982) 250. 
%
%
%
%\bibitem{STT1}
%K. Shima, Y. Tanii and M. Tsuda, {\it Phys. Lett. B} {\bf 525} (2002) 183. 
%
%\bibitem{STT2}
%K. Shima, Y. Tanii and M. Tsuda, {\it Phys. Lett. B} {\bf 546} (2002) 162. 
%
%\bibitem{ST1-lin}
%K. Shima and M. Tsuda, {\it Phys. Lett. B} {\bf 641} (2006) 101. %($N = 3$, $d = 2$, free)
%
%\bibitem{ST2-lin}
%K. Shima and M. Tsuda, {\it Mod. Phys. Lett. A} {\bf 23} (2008) 3149. %($N = 2$, $d = 2$, free, SF)
%
%%%%%%%%%%%%%%%%%%%%(NL/L SUSY relation for interacting theories)
%
%\bibitem{ST3-lin}
%K. Shima and M. Tsuda, {\it Mod. Phys. Lett. A} {\bf 22} (2007) 1085 %($N = 2$, $d = 2$, Yukawa and mass terms). 
%
%\bibitem{ST4-lin}
%K. Shima and M. Tsuda, {\it Phys. Lett. B} {\bf 666} (2008) 410. %($N = 2$, $d = 2$, WZ-gauge, vector SF)
%
%\bibitem{ST5-lin}
%K. Shima and M. Tsuda, {\it Mod. Phys. Lett. A} {\bf 22} (2007) 3027. %($N = 2$, $d = 2$, QED)
%
%\bibitem{ST6-lin}
%K. Shima and M. Tsuda, {\it Mod. Phys. Lett. A} {\bf 24} (2009) 185. %($N = 2$, $d = 2$, QED, scalar and vector SF)
%
%\bibitem{ST7-lin}
%K. Shima and M. Tsuda, {\it Phys. Lett. B} {\bf 687} (2010) 89. %($N = 2$, $d = 2$, super YM)
%
%%%%%%%%%%%%%%%%%%%%
%
%\bibitem{ST-vac}
%K. Shima and M. Tsuda, {\it Phys. Lett. B} {\bf 645} (2007) 455. 
%
%\bibitem{STL}
%K. Shima, M. Tsuda and W. Lang, {\it Phys. Lett. B} {\bf 659} (2008) 741. 
%
%\bibitem{BV}
%W. Bardeen and V. Vi\v{s}nji\'{c}, {\it Nucl. Phys. B} {\bf 194} (1982) 422. 
%
%\bibitem{WB}
%J. Wess and J. Bagger, {\it Supersymmetry and Supergravity 
%(Second Edition, Revised and Expanded)} 
%(Princeton University Press, Princeton, New Jersey, 1992). 
%
%\bibitem{DVF}
%P. Di Vecchia and S. Ferrara, {\it Nucl. Phys. B} {\bf 130} (1977) 93. 
%
%\bibitem{ST-SF}
%K. Shima and M. Tsuda, {\it Mod. Phys. Lett. A} {\bf 23} (2008) 1167. 
%
%\bibitem{ST-gauge}
%K. Shima and M. Tsuda, {\em Nuovo Cimento} {\bf 124B} (2009) 645. 
%
%\bibitem{CEKKLW}
%D.J.H. Chung, L.L. Everett, G.L. Kane, S.F. King, J. Lykken and Lian-Tao Wang, 
%{\it Phys. Rept.} {\bf 407} (2005) 1. 
%
%\bibitem{FI}
%P. Fayet and J. Iliopoulos, {\it Phys. Lett. B} {\bf 51} (1974) 461. 

\end{thebibliography}
\end{document}